\documentclass[a4paper]{article} 
\usepackage{t1enc}
\usepackage[latin2]{inputenc} 
\usepackage{amsmath}
\usepackage{amssymb} 
\usepackage{graphicx} 
\usepackage{subfigure}

\usepackage{times}

\begin{document}

\title{Hierarchically Organized Minority Games}

\author{Csaba F\"oldy \dag,  \hspace*{0.1cm} Zolt\'an Somogyv\'ari \dag,
\hspace*{0.1cm} P\'eter \'Erdi  \ddag \dag \\
\dag Department of Biophysics, KFKI  Research  Institute for Particle \\
and Nuclear Physics of the Hungarian Academy of Sciences \\ H-1525 
Budapest  P.O.  Box 49, Hungary \\
\ddag Center for Complex System Studies, \\ Kalamazoo College, Kalamazoo,
MI 49006 , U.S.A}

\maketitle
 
\begin{abstract} In this paper  a hierarchical extension  of the Minority
Game is defined and studied.   Numerical simulations show a special type
of emergent global behavior between separated parts of the hierarchical
structure, connected only through a normalized mean field quantity. \\

\end{abstract}

\section{INTRODUCTION}

In this paper a simple, single-level strategic game, namely the Minority
Game (MG), will be extended to form a hierarchical  organization. The
individual MGs will form a  group, and individual players within each MG
will make decisions which depend on the average behavior of their own and 
other MGs. Calculations show  that under certain conditions the elements of
the individual MGs ``know'' much more about each other than one would
expect.
\\

"...Hierarchical organization is a common way to structure a group of
people, where members chiefly communicate  with their immediate superior
and with their immediate subordinates. Structuring organizations in this
way is useful partly because it can reduce the communication overhead."
\cite{wikipedia}. \\

One of the most interesting questions in the theory of hierarchical
systems is how much information  a subsystem has about the performance of
the whole.  In addition to study this problem the top-down influence of
the "whole" system to  the elements  will also be quantitatively 
studied.  \\

\section{HIERARCHICAL EXTENSION of MG}

MG is a simple model of inductively rational systems \cite{challet97}.
This game is a system of $N$ interacting agents. Each agent brings a
decision $a_i\left(s_i(t),\mu(t)\right)$ between $+1$ and $-1$  based on
her $s_i(t)$ strategy and the knowledge of the history of the game
$\mu(t)$. In all simulations each agent chooses from one of two
stategies which are randomly chosen for each agent in the beginning of the
game. A strategy is a
look-up table which assigns the actual choice to the M component vector
$\mu(t)$ of the previous M outcome signs, $sgn(A(t))$. $A(t)$ is defined
as
$A(t)=\sum_ia_{i}\left(s_{i}(t),\mu(t)\right)$. Each agent chooses the
$s^{th}$ strategy from the possible ones with probability

\begin{equation}
Prob\{s_{i}\left(t\right)=s\}=Z_{i}e^{\Gamma_{i}U_{is}\left(t\right)}
\end{equation} Where
$Z_{i}^{-1}=\sum_{s'}e^{\Gamma_{i}U_{is'}\left(t\right)}$, $\Gamma_{i}$
is the inverse temperature, and the performance of the $i^{th}$ agent's
$s^{th}$ strategy is evaluated by a cumulative score $U_{is}(t)$. The
updating rule for the evaluation:
\begin{equation}
U_{is}\left(t+1\right)=U_{is}\left(t\right)-a_{i}\left(s_{i}(t),\mu(t)\right)A(t)
\end{equation} The second term represents the gain of the individual
agents.\\

 The original model is extended to form a hierarchical structure. This
extension is done in the following way. $N'$  original MGs
are connected into a new MG, where the original MGs are the agents of
this new game. In this hierarchical organization there are $N'$
($j=1...N'$) original MGs, in each of which $N$ ($i=1...N$)
(low level) agents play and where their decision is $a_{ij}(s_{ij}(t),\mu_j(t))$. We use $\mu_j$ since each local MG's history contains the local and global minority sign hence it is different for every local MG. For simplicity we will denote 
$a_{ij}\left(s_i(t),\mu_j(t)\right)$ by $a_{ij}\left(t\right)$. \\

In each low level MG we have an outcome minority sign,
$-sgn(A_j)$. From the set of these signs a global sum and a minority sign,
 is defined as $A=\sum_{j}sgn(A_{j})$ and $-sgn(A)$. Thus the
score updating dynamics of the strategies are modified as: 
\begin{equation} U_{ijs} (t+1)=U_{ijs}(t)- a_{ij}(t) \left( A_j(t)  + C
\sum_k sgn\left(A_k (t)\right) \right)\label{e1},
\end{equation}

The constant, $C$,  expresses the ratio of the contribution of the  local
and global environments.  The local environment consists of the agents
playing in the original MG, while   the global one is derived from the
connection of the MGs. The game among the MGs, however,  differs from the
original. The  MGs, as individual agents, don't have strategies and
memories explicitly. Instead the global connection between the MGs is made
only through their low level agents.

\begin{center}
\begin{figure*}[here!]
\includegraphics*[width=120mm, height=100mm]{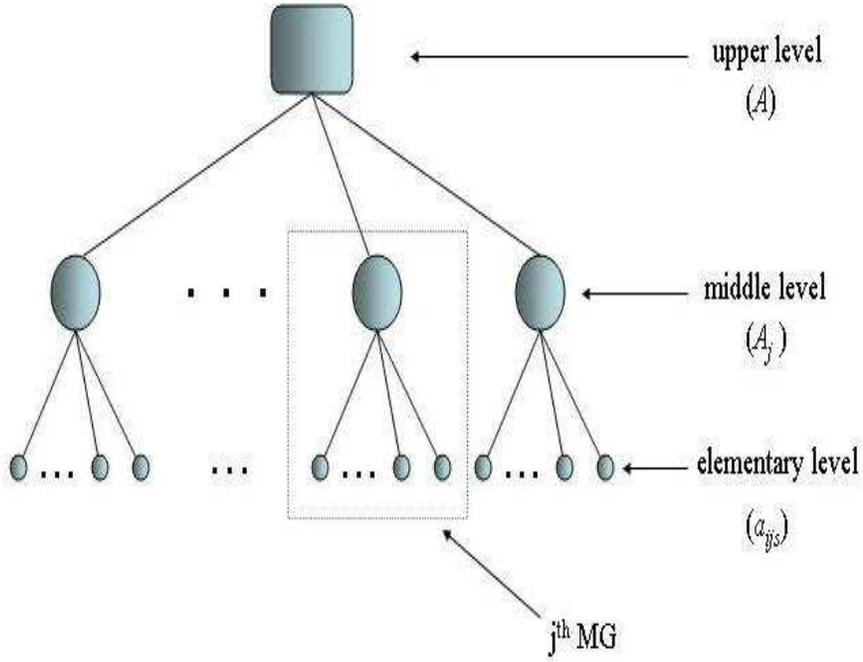}
\caption{Hierarchical organization of the minority games}
\end{figure*}
\end{center}

When $C \rightarrow 0$ we obtain the original definition. It is often
assumed in  the theory of hierarchical  structures  that the  local
interactions should be much stronger than the global ones
\cite{auger89}.  However, the violation of this assumption implies
non-trivial results. \\

In the context of MG for values of $C$ when the global interactions are
stronger than the local ones, the  agents   don't choose a strategy to
maximize their own winning chance only, but also to help  their  whole
group win against other groups. \\

To be able to analyze our extended multilevel model, we will decompose it
into three subsystems.  Each subsystem  contains binary interlayer
interactions only, as shown in Fig.(1). The characteristic quantities,
namely the global efficiency  and the
degree of symmetry, will be properly redefined.

\section{RESULTS}

\subsection{Interactions between the elementary and intermediate levels}

The new additive part of Eq.(\ref{e1})
is the same for every elementary  participant, therefore we  expect that
the normalized global ``waste'' $\sigma_j^2 / N$ averaged over the $N'$
MG's (denoted by $<\sigma_j^2 / N>_{N'}$)
will show the same characteristic pattern as in the  original game. 
A small value of the waste implis a high value of efficiency.

By this extension,
however,  the total payoff of the
$j^{th}$ MG 
$u_j = -A_j^2 - A_j A$ might  be greater than zero, and thus it is no
longer a "negative sum game". 

\begin{center}
\begin{figure*}[here!]
\includegraphics*[width=40mm, height=65mm, angle=-90]{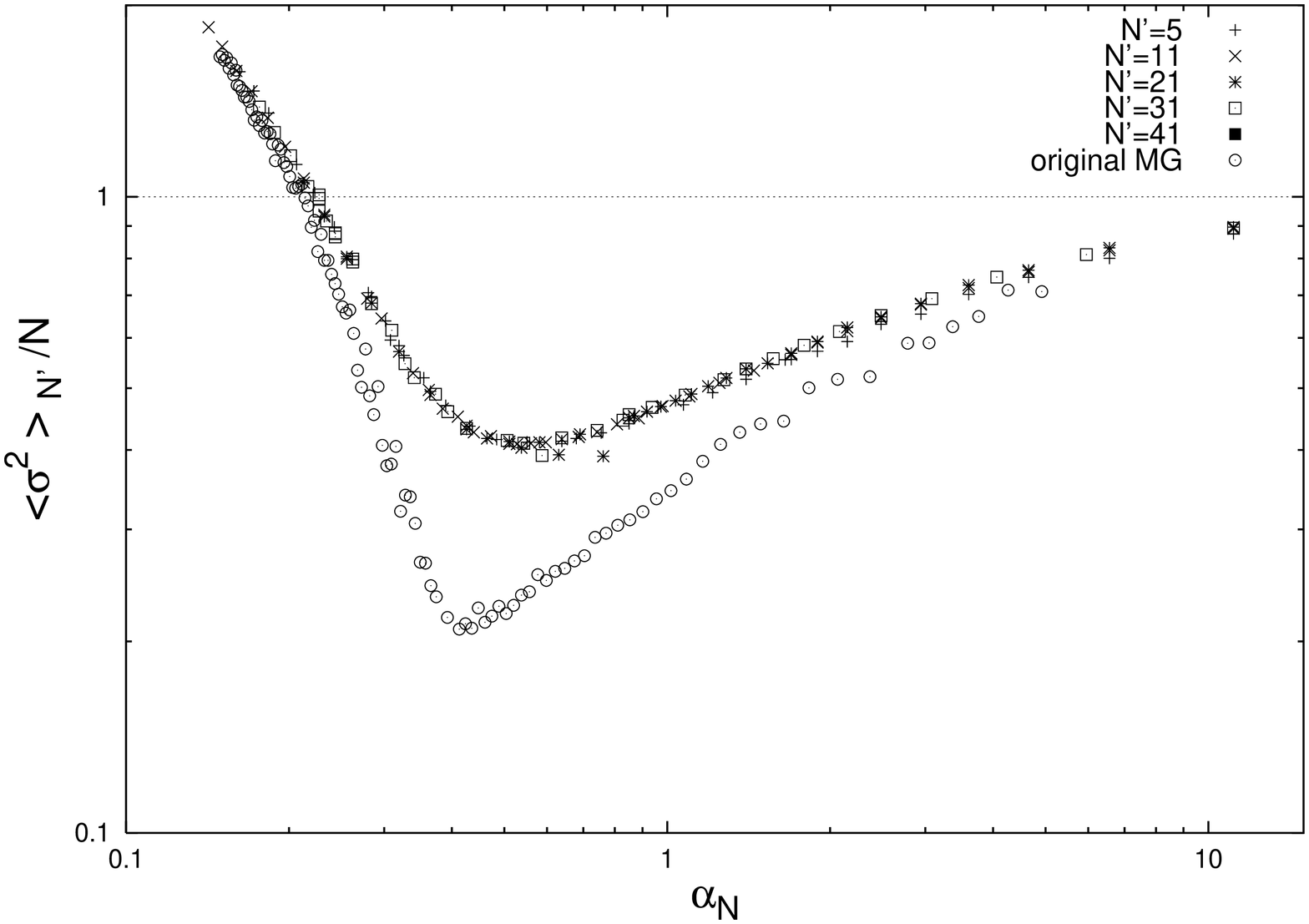}
\includegraphics*[width=40mm, height=65mm, angle=-90]{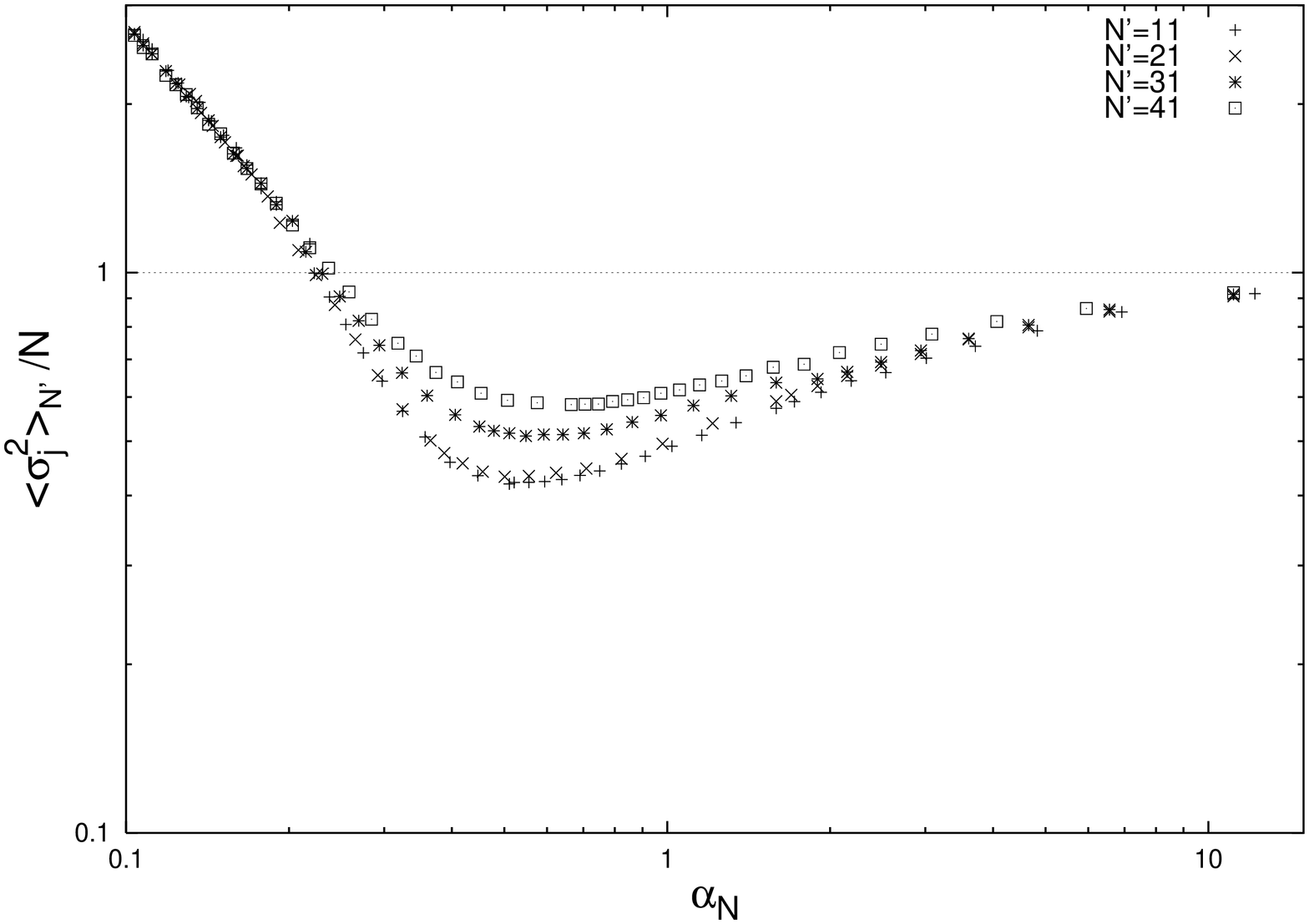}
\caption{$<\sigma_j^2 /N>_{N'}$ as a function of $\alpha_N = P_{em}/N$ for various
$N'$s, with
$C=1/N'$ (left) and $C=1$ (right).$\Gamma_{ij}
\sim \infty$, $U_{ijs}(0)=0$, for $
\forall$ $i,j,s$.} \label{abr2}
\end{figure*}
\end{center}

Fig.(2) shows the results of the comparative studies between the
original and the special features of the
 extended game by calculating $\sigma_j^2$ using the equation
\begin{equation}
\sigma_j^2 \equiv \overline{\langle A_j^2 \rangle}=\sum_{\mu
_j=1}^{P_{em}}\rho^{\mu_j}
\left\langle \left(\sum_{i=1}^{N} - a_{ij}(t) A_j^{\mu_j}(t)\right)^2 \right\rangle
\end{equation}
where $P_{em}=2^M$ is the number of possible memory states, $\rho^{\mu_j}$ is the
probability of occurrence of the $\mu$th state in the $j$th MG. The $\left<...\right>$
refers to the averaging over the possible strategies and the overline
stands for the averages over the memory states. \\       

The left part of Fig. \ref{abr2} refers to the case of $C=1/N'$ for
various values of $N'$. Simulations show that  the average of $\sigma_j^2 /
N$  deviates from that obtained in the  original game, but the deviation
does not depend on the value of $N'$.  In the right part of Fig.(2)
($C=1$), the average of $\sigma_j^2 / N$ differs not only from the
original characteristics, but is different for each value of $N'$. It is
easy to understand this dependence on  $N'$, because the additive part in
Eq.(\ref{e1}) does not implement any normalization. (The order of
magnitude of the new additive factor is  $o (N N')$, while the original
part of Eq.(\ref{e1}) is just
$o (N)$. For larger values of our new additive part (which is 
proportional to $N'$) and by the definition of MG we obtain larger values
of
$\sigma_j^ 2 / N$ than one would see in the original game.  \\

\begin{center}
\begin{figure*}[here!]
\subfigure{\includegraphics*[width=40mm, height=65mm,angle=-90]{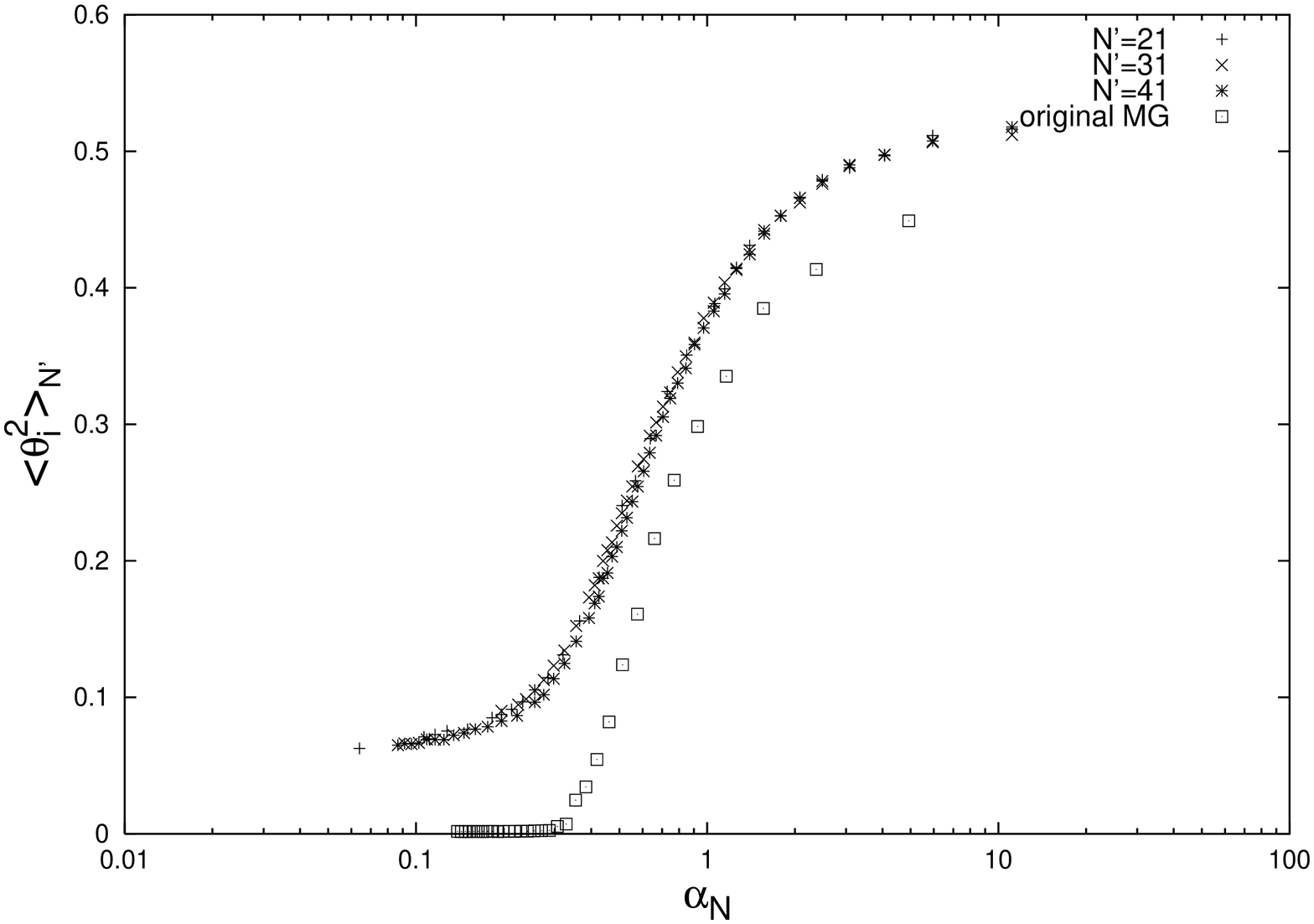}}\qquad
\subfigure{\includegraphics*[width=40mm, height=65mm, angle=-90]{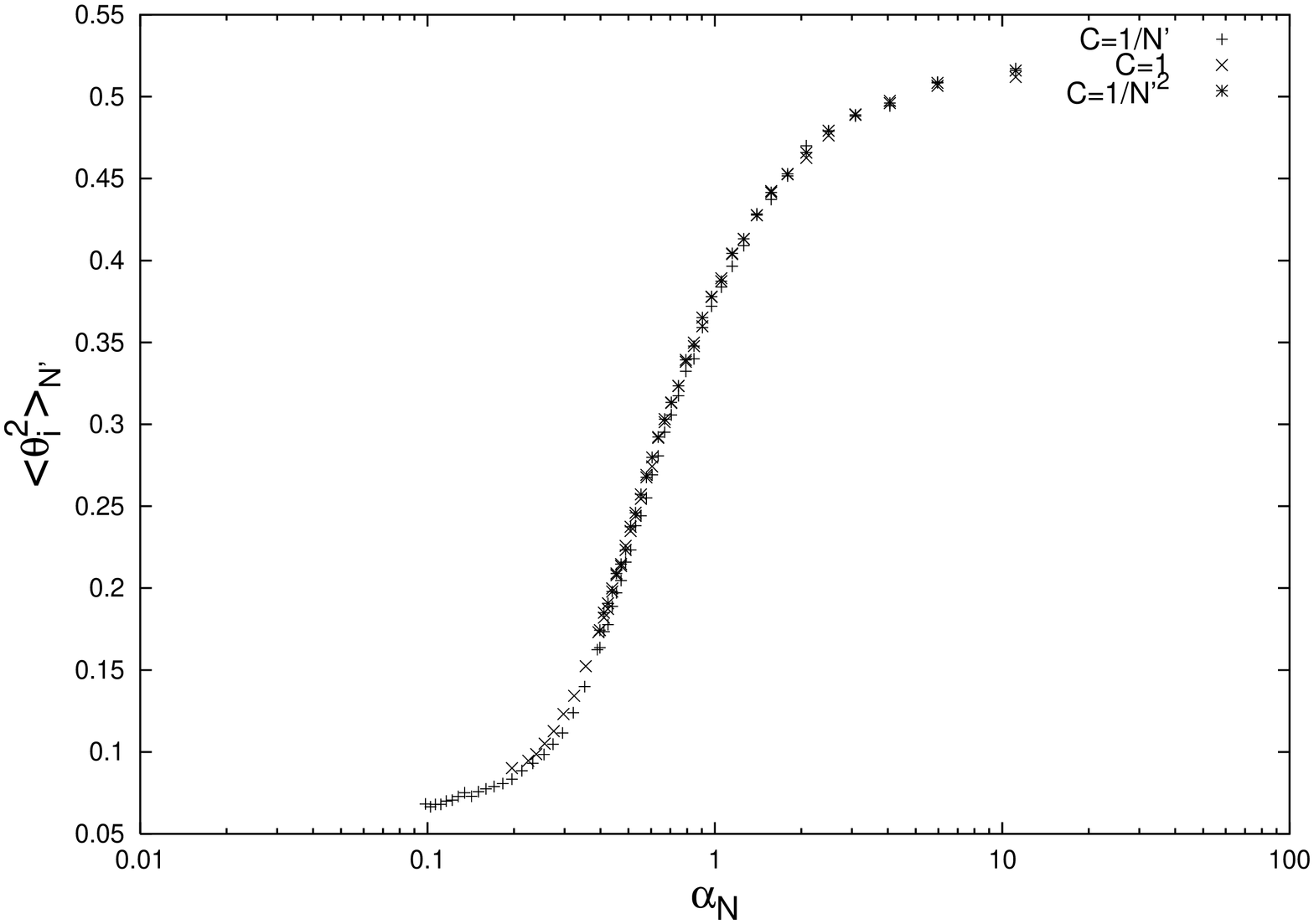}}
\caption{$<\theta_j^2>_{N'}$  as a function of $\alpha_N$  for various $N'$s (left) and
$C$s (right). $\Gamma_{ij}
\sim \infty$, $U_{ijs}(0)=0$, for $ \forall$ $i,j,s$.} \label{abr3}
\end{figure*}
\end{center}

A more striking result was obtained for the behavior of 
the symmetry between the two  minority signs. Fig.(\ref{abr3}) plots the
degree of asymmetry ($\theta_j^2$) as a function of $\alpha_{N}$ (left)
and $C$ (right), respectively. The asymmetry is defined by
\begin{equation}
\theta_j=\sqrt{\frac{1}{P_{em}}\sum_{\mu_j=1}^{P_{em}}\left<-sgn(A_j(t))\vert\mu_j\right>^2}
\end{equation}
where $\left<-sgn(A_j(t))\vert\mu_j\right>$ is the conditional probability
of $-sgn(A_j(t))$ conditioned on $\mu_j$. \\

 Our numerical results show that the system nowhere became symmetrical
for the two signs. Indeed in Fig.(\ref{abr3}) we can see that for various
$C$s and $N'$s we obtain the same characteristics reflecting  that there
is no phase transition in the system with the parameters we investigated.
The tendency of the change  occurs more slowly  than in the original
game. The difference appears because our subsystems are not closed
because of the interaction between the MGs. Though there exists an
`optimal strategy' where the symmetrical phase should appear, it's 
efficiency is very low.   

\subsection{Interactions between the middle and upper levels}

Our naive expectation is to find 'random players,' i.e. players who
select  their strategies randomly, because the choices of the middle
level players are just  the results of minority games of the lower level.
These 'choices' cannot be interpreted as the results  of 'decisions'
because the middle level does not have any direct memories or
strategies.  They are only mean field averages over the individual MGs.
Although the lack of memory implies the break down of the definition of
the original scaled parameter, $\alpha_N$,  we still found this quantity
to be useful. \\

We define the waste $\sigma_{mu}^2$ in terms of the variation in each of
the middle level games ($mu$ is the symbolic notation of ''between middle and upper''). Explicitly, we have
\begin{equation}
\sigma_{mu}^2 \equiv \overline{\langle A^2 \rangle}=\sum_{j=1}^{N'}\sum_{\mu_j
=1}^{P_{em}}\rho^{\mu_j}
\left\langle \left( \sum_{j=1}^{N'} -sgn\left( \sum_{i=1}^{N} a_{ij}(t) A_i^{\mu_j}(t)\right)A(t)\right)^2 \right\rangle  
\end{equation} 
Note that the last term of Eq.(6) depends on $N'$. The lack of normalization will not have any effect, 
because we are only interested in the sign of the sum of the local agents' performance. 

\begin{center}
\begin{figure*}[here!]
\subfigure{\includegraphics*[width=40mm, height=65mm, angle=-90]{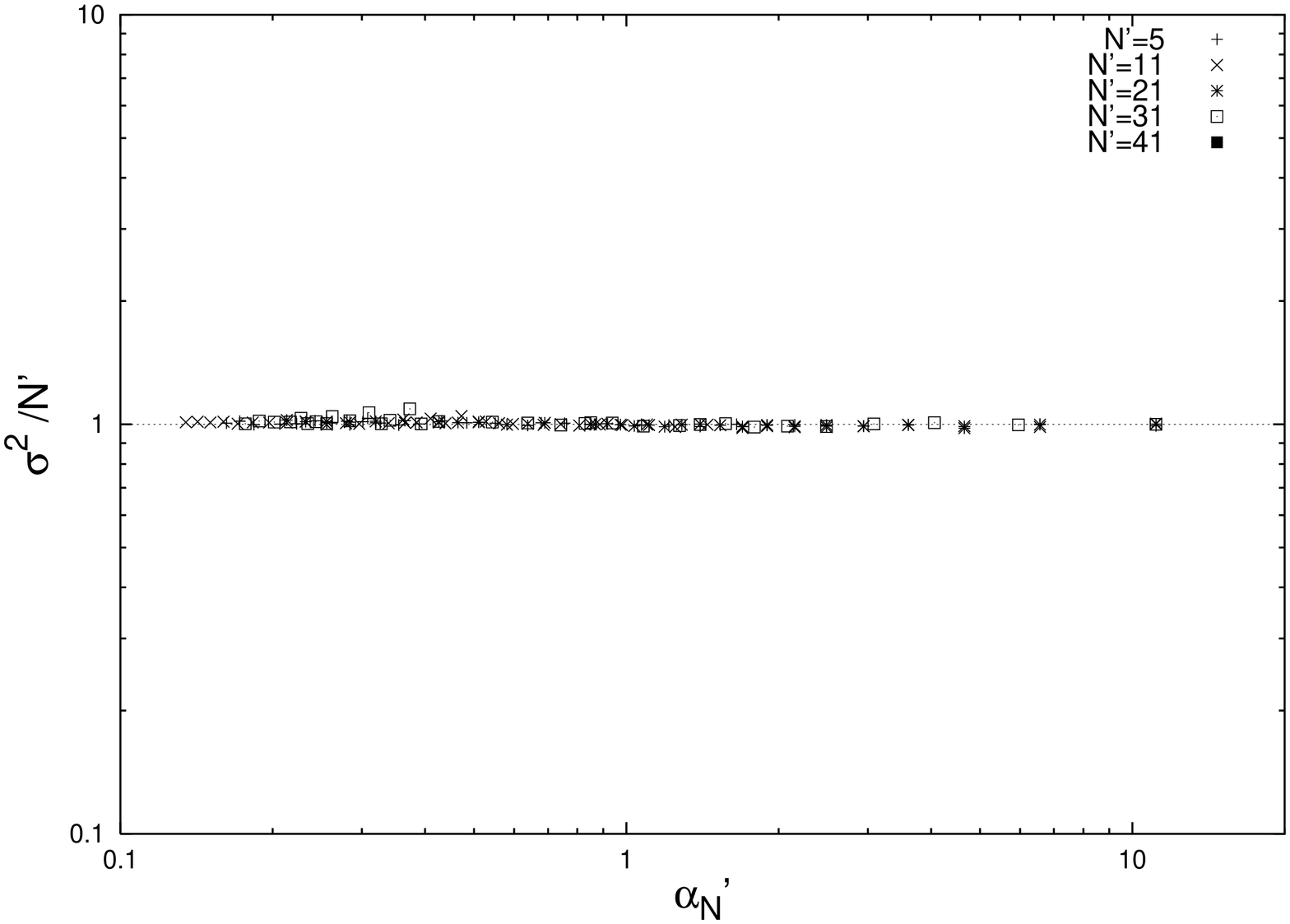}}
\subfigure{\includegraphics*[width=40mm, height=65mm,
angle=-90]{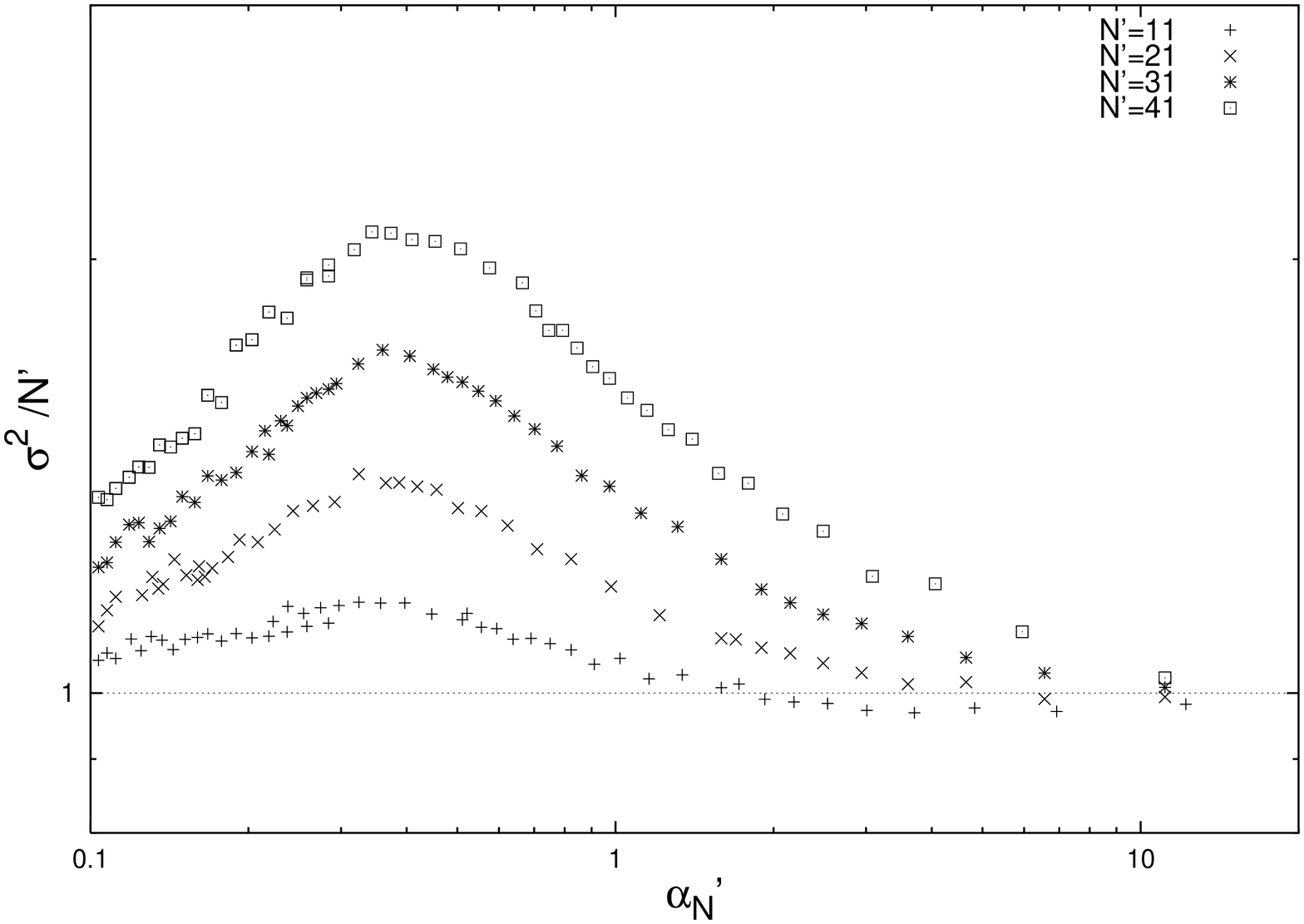}}
\caption{$\sigma_{mu}^2 / N'$ as a function of $\alpha_N' = P/N'$ for
various
$N'$s, with
$C=1/N'$ (left) and $C=1$ (right). 
$\Gamma_{ij} \sim \infty$, $U_{ijs}(0)=0$,  for $
\forall$ $i,j,s$.}\label{abr4}
\end{figure*}
\end{center}

The simulation results reflect the expectations discussed above. For
those values of $C$ which implement larger local  connections within an 
MG than to the global one ($C=1/N'$) the system shows
random behavior (Fig.\ref{abr4} left). When the global connection gets
stronger, i.e. $C=1$, the system shows an emergent  coordination
\cite{bottazzi}, i.e. the system as a whole shows a different than random
performance. Fig.(\ref{abr4}) (right) shows positive and not negative
deviation from the random performance. This phenomenon  is the result of
the interaction of the two minority games.

\subsection{Interactions between the elementary and upper levels}

In this case we could ask whether there is any direct connection between
the elementary and upper levels. Here we calculated different globally
averaged functions
$\sigma_{eu}$ and
$\theta_{eu}^2$ (where $eu$ the symbolic notation of ''between elementary and upper''). We define 
\begin{equation}
\sigma_{eu}^2 \equiv \overline{\langle A^2 \rangle}=\sum_{\mu
=1}^{P_{eu}}\rho^{\mu}
\left\langle \left( \sum_{j=1}^{N'} \sum_{i=1}^{N} a_{ij}(t) A^{\mu}(t)             
\right)^2 \right\rangle  
\end{equation} 
First, we note that there is
now another scaled parameter.  For 
$P_{eu}=(2^{\frac M2})^{N'+1}$ possible states of the system the new scaled
parameter is
$\alpha_{N,N'}=\frac{P_{eu}}{NN'}$. And $\mu$ refers to the state of the whole system (i.e. $\rho^{\mu} \in \bigotimes_{j=1}^{N'} \bigoplus_{\mu_j=1}^{P_{em}} \rho^{\mu_j}$).

\begin{center}
\begin{figure*}[here!]
\includegraphics*[width=40mm, height=65mm, angle=-90]{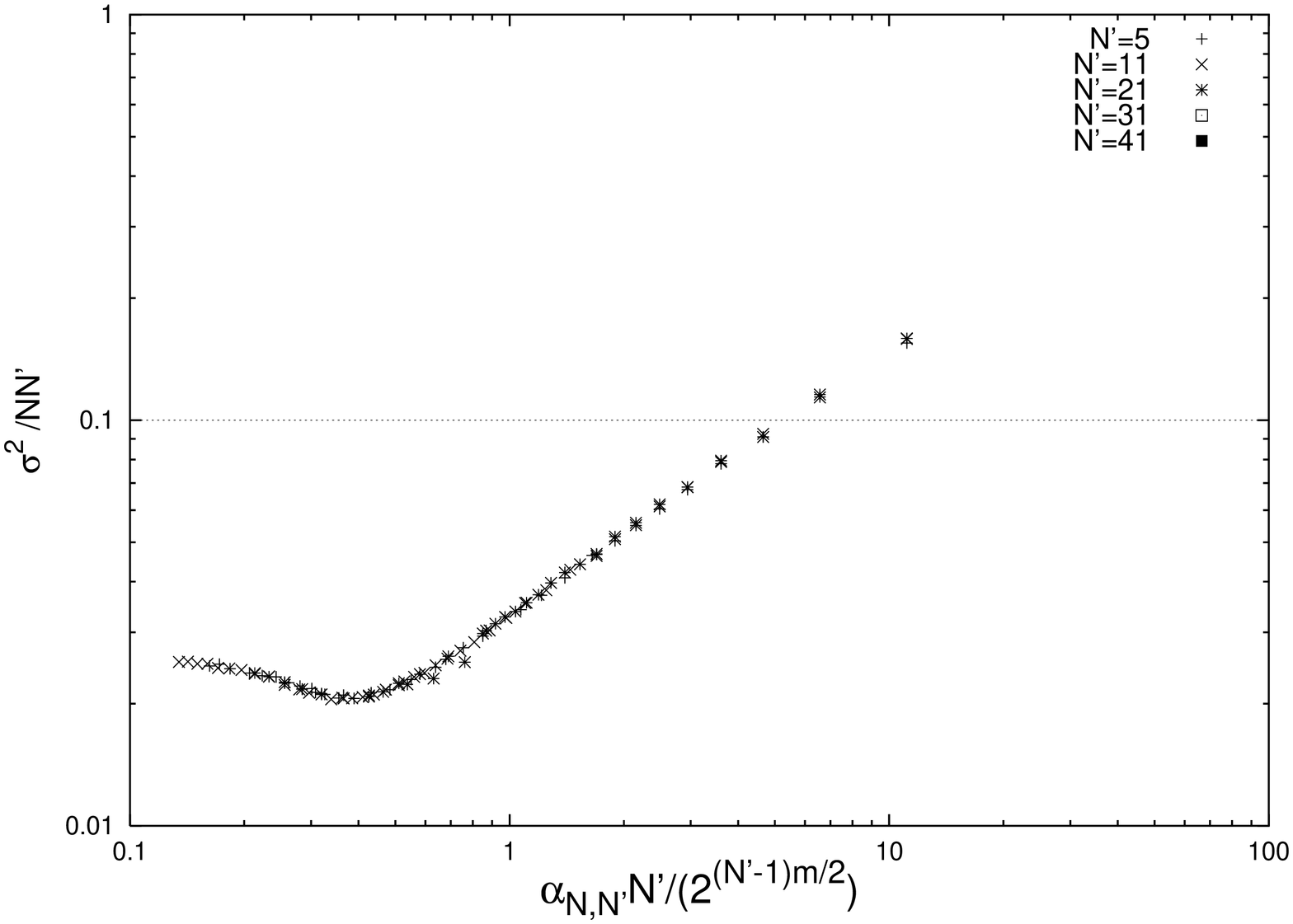}
\includegraphics*[width=40mm, height=65mm, angle=-90]{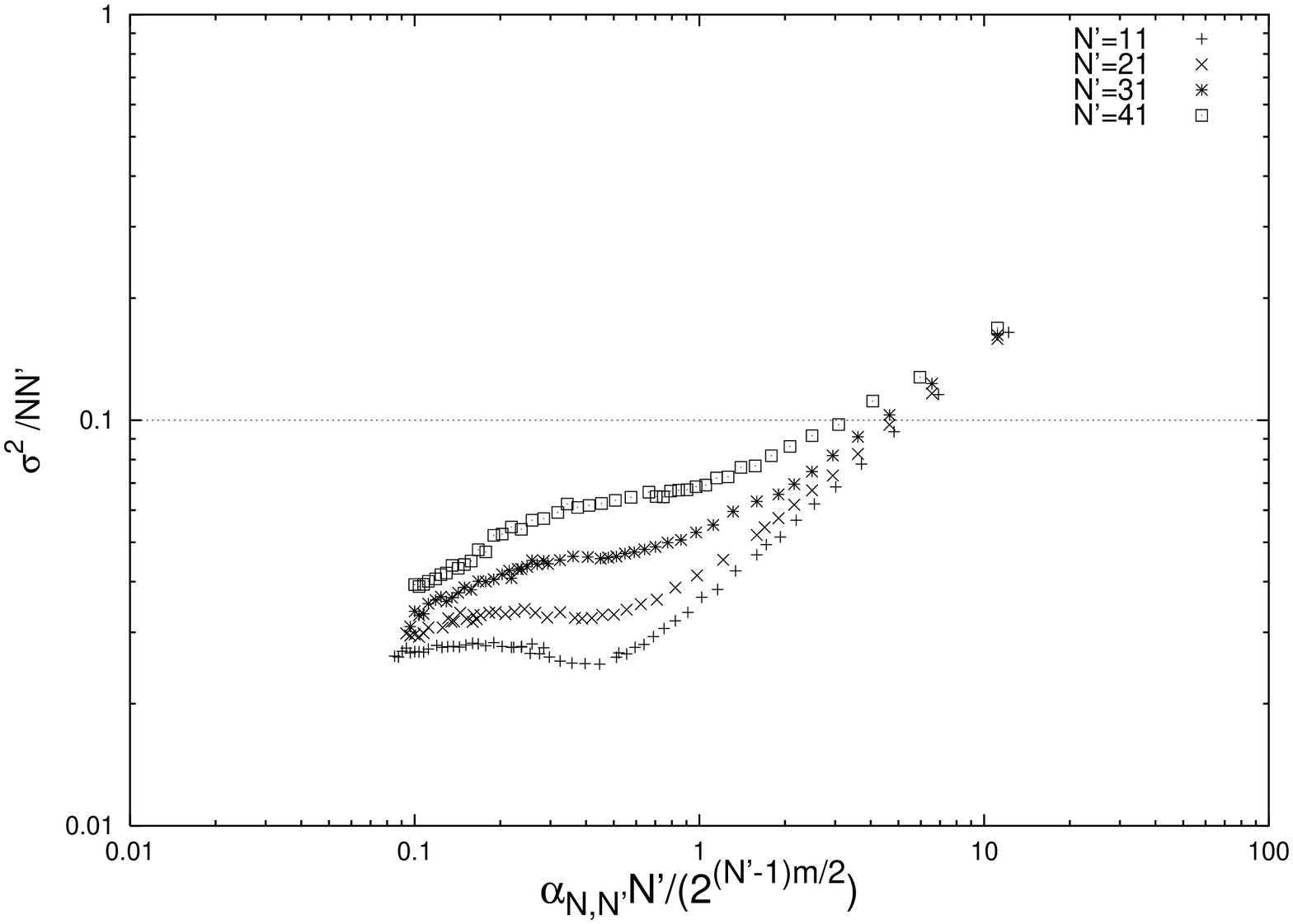}
\caption{$\sigma_{eu}^2 / (NN')$ as a function of $\alpha_{N,N'}$ for
various
$N'$s, with
$C=1/N$ (left) and $C=1$ (right). 
$\Gamma_{ij} \sim \infty$, $U_{ijs}(0)=0$, for $ \forall$
$i,j$.}\label{abr5}
\end{figure*}
\end{center}

In these numerical simulations (Fig.(\ref{abr5})) the results are
qualitatively the same. (Without normalization  the value of
$\sigma_{eu}^2/(NN')$ increases for greater values of $N'$.) \\

\begin{center}
\begin{figure*}[here!]
\includegraphics*[width=40mm, height=65mm, angle=-90]{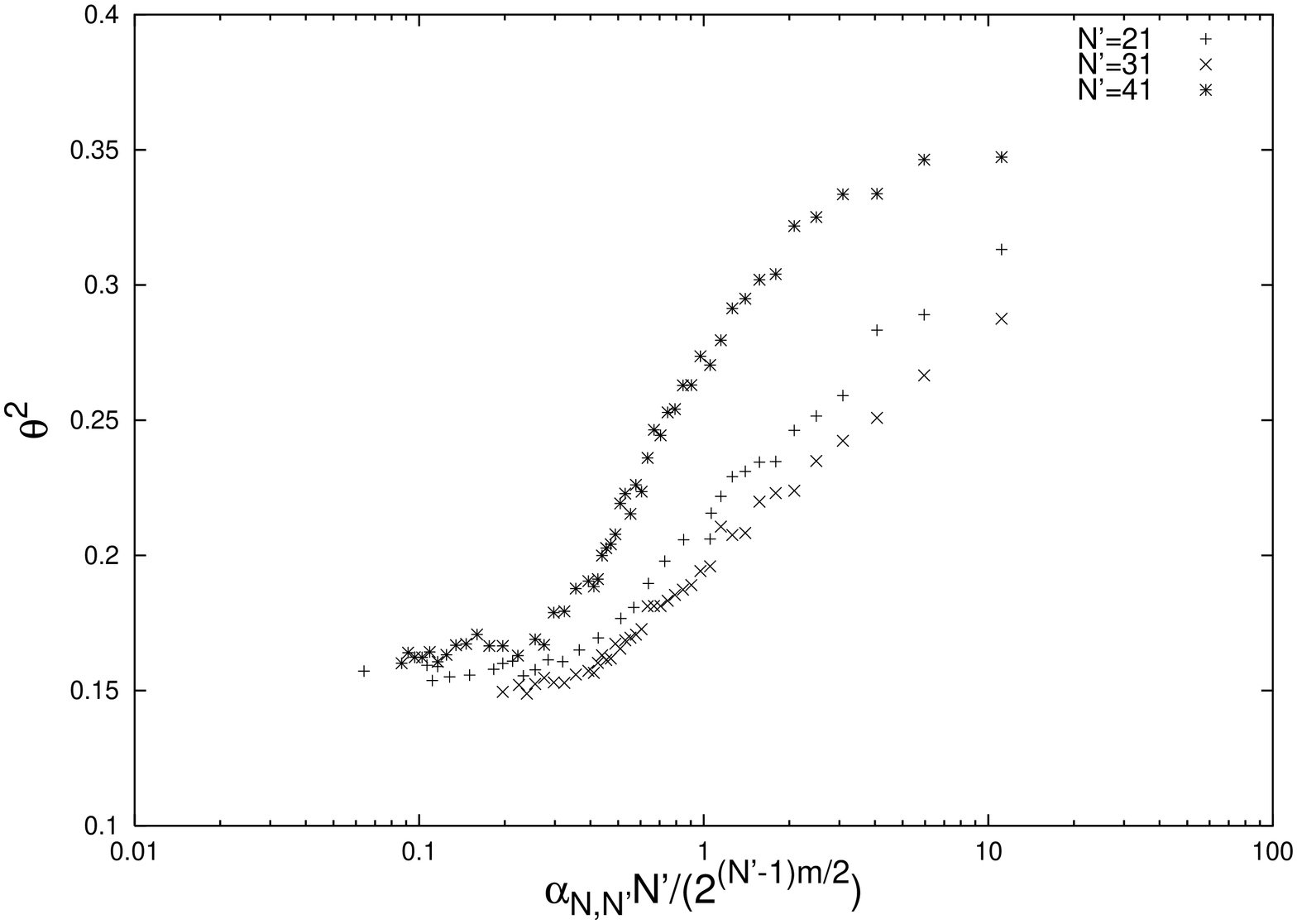}
\includegraphics*[width=40mm, height=65mm, angle=-90]{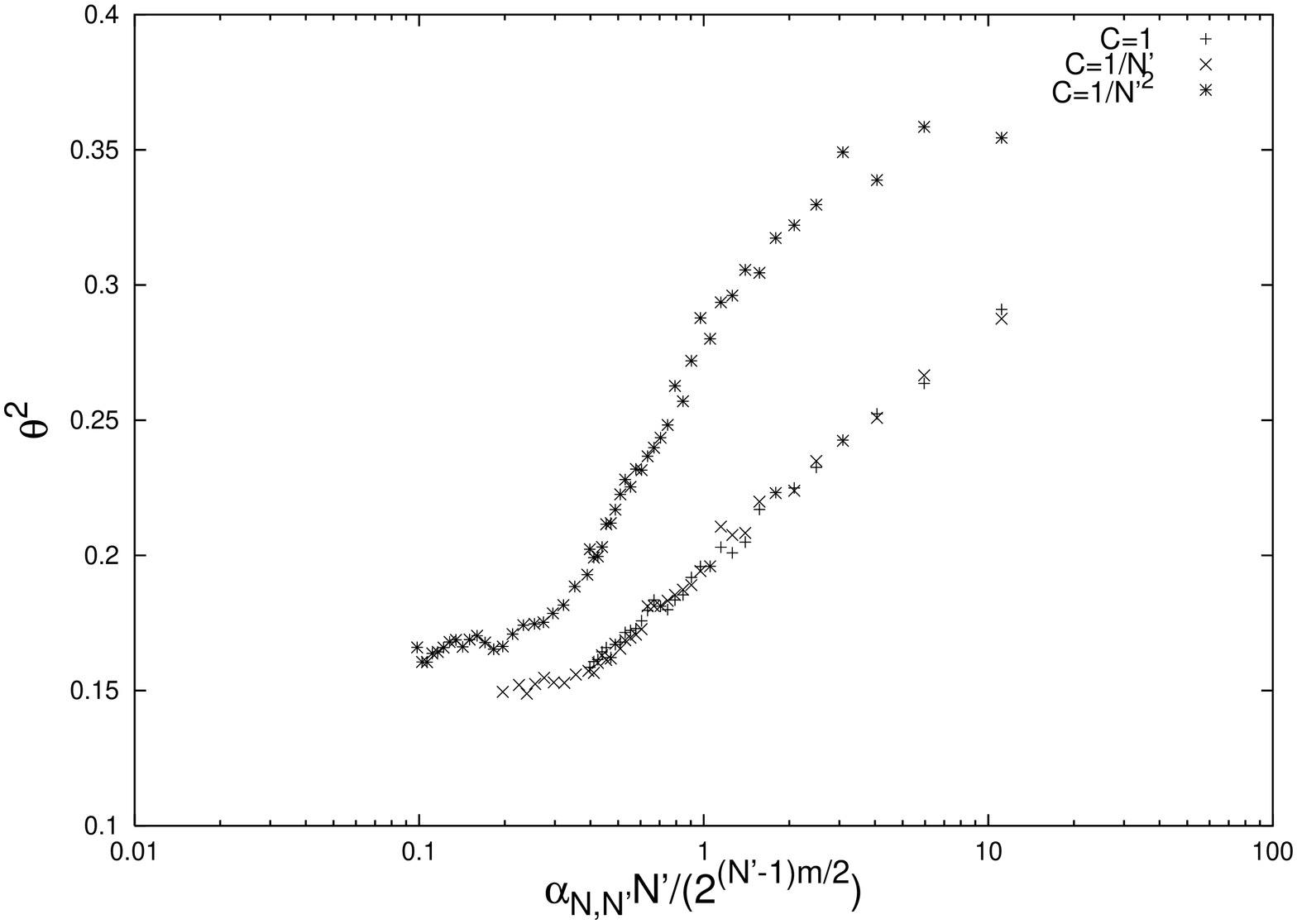}
\caption{$\theta_{eu}^2$ as a function of $\alpha_{N,N'}$ for various
values of $N'$ (left) and
$C$ (right).
$\Gamma_{ij} \sim \infty$, $U_{ijs}(0)=0$, for $ \forall$
$i,j,s$.}\label{abr6}
\end{figure*}
\end{center}

Another interesting result of our investigation is the symmetry at this
level. One might believe  that  $\theta_{eu}^2$  would not depend on
either $N'$ or $C$, explicitly. However, the simulation result
(Fig.\ref{abr6}) still shows some dependence on $N'$ and $C$. 

\

\section{Discussion}

We studied a hierarchical extension of the MG for three levels and
studied the inter-level interactions.  First, the local MGs, i.e. the
connections between the lower and middle layers were studied. It was
interesting  to see that for the ratio $C=1$ the global efficiency
depends on the number of individual MGs.  By using this formula the
individual agents know the number of all MGs. In addition, at least in a
broad region  of the parameters, the degree of symmetry is never  zero.
Consequently, in contrast to the original game,  there always exists  an
optimal strategy for winning the game. \\

Second, the connections between the intermediate and the upper levels
were investigated. For the case of relatively strong local connections,
the system seems to behave randomly. For strong global coupling the
behavior deviates from  random behavior, and shows a  self-organized
global behavior, which can be considered a special type  of 'emergent
coordination.' \\

Finally, the results of the study of the connections between the lower
and the upper level also exhibits the  property seen previously: for the
case of $C=1$ the global efficiency depends on the number of games.  In
addition, the global efficiency shows interesting dependences both on the
values of the ratio number, and  on the number of games. In the future we
intend to investigate these dependences analytically. \\

\section{Acknowledgments} We thank Jan Tobochnik for carefully reading
our manuscript and providing helpful suggestions.  This research was
supported by  OTKA T038140, and AKP 2000-148 2,3. Special  thanks to the
Henry Luce Foundation for support.

\end{document}